\documentclass[lettersize,journal]{IEEEtran}
\usepackage{amsmath,amsfonts}
\usepackage{algorithmic}
\usepackage{algorithm}
\usepackage{array}
\usepackage[caption=false,font=normalsize,labelfont=sf,textfont=sf]{subfig}
\usepackage{textcomp}
\usepackage{stfloats}
\usepackage{url}
\usepackage{verbatim}
\usepackage{graphicx}
\usepackage{cite}
\begin{document}

\title{Shot-noise limited, 10 MHz swept-source optical coherence tomography for retinal imaging}

\author{Sacha Grelet, Alejandro Martinez Jimenez, Patrick B. Montague, Adrian Podoleanu
    \thanks{Sacha Grelet, Alejandro Martinez Jimenez, and Adrian Podoleanu are with the Applied Optics Group, School of Physical Sciences, University of Kent, Canterbury CT2 7NH, United Kingdom.}

    \thanks{Sacha Grelet and Patrick Bowen Montague are with NKT Photonics A/S, Blokken 84, DK-3460, Birkerød, Denmark.}}

\markboth{October~2024}%
{Shell \MakeLowercase{\textit{et al.}}: A Sample Article Using IEEEtran.cls for IEEE Journals}


\maketitle

\begin{abstract}
Akinetic swept-sources are essential for high-speed optical coherence tomography (OCT) imaging. Time-stretched supercontinuum (TSSC) lasers have proven to be efficient for multi-MHz swept-sources. However, lack of low-noise broadband lasers and of large dispersion devices in the water low-absorption band at 1060\,nm have limited the biomedical applications of TSSC lasers. In this letter, an approach to tune the wavelength around 1050\,nm over 90\,nm with low-noise at 10\,MHz is presented. This is based on all-normal dispersion (ANDi) supercontinuum dynamics, and employs a long chirped fiber Bragg grating (CFBG) to time-stretch a broadband pulse with a duty cycle of 93\%. Retinal images are demonstrated, with a sensitivity of 84\,dB - approaching the shot noise limit. We believe this high-speed low-noise swept-source will greatly promote the development of OCT techniques for biomedical applications.
\end{abstract}

\section{Introduction}
\IEEEPARstart{R}{etinal} optical coherence tomography (OCT) imaging is crucial for identifying diseases such a glaucoma and age-related macular degeneration (AMD) \cite{bussel2014oct, schlegl2018fully}. There is an increasing demand for high acquisition speed OCT to improve the clinical throughput, to overcome motion artifacts from moving samples, and to obtain additional information such as that provided by OCT angiography (OCTA) \cite{Draxinger:24, Wei2019,Kim2020}. Swept-source OCT (SS-OCT) is the most promising technology for achieving high-speed flying spot imaging, already demonstrating MHz A-scan rate \cite{Klein2017}.

The major challenge of flying spot SS-OCT resides in the light source. The most widely used operations principles employ micro-electromechanical system (MEMS) vertical cavity surface emitting laser (VCSEL) \cite{Potsaid2012}, or Fabry-Pérot tunable filters \cite{huber2005amplified}. Due to mechanical constrains, their sweeping-rate is limited to the MHz range and their tuning curve is nonlinear in wavenumber, necessitating additional signal processing steps. Higher sweep rates have been achieved with high-order buffering, which multiplies the sweep rate but also increases assembly complexity, posing challenges for manufacturing at scale.

For further increase of the sweeping rate, akinetic swept-sources have been developed, based on dispersion for broadband time-stretching. One version of this implementation, stretch-pulse modelocking (SPML), places the stretcher inside the cavity of an active mode-locked laser \cite{Kim2020,Akkaya2022}. While promising, SPML lasers require an expensive intensity modulator as well as triggering of extreme precision. 

A second version of the time-stretching is to use a stretcher outside the laser cavity. The sweeping rate of such systems is determined by the repetition rate of the laser oscillator, which can easily be designed between 10-100\,MHz. This offers more freedom in the design of the swept-source, as the sweeping bandwidth (laser bandwidth) and the sweeping rate (dispersive stretching) are decoupled. Using this method, 100-400\,MHz swept-sources have been demonstrated \cite{Huo2015,Huang2020}. However, existing publications employ lasers with a spectrum centered at 1550\,nm, which is not practicable for biomedical imaging due to the considerable water absorption. For this reason, it is of interest to develop such system at 1060\,nm where the local minimum of the water absorption spectrum already attracted attention in imaging the retina in the human eye.

Adapting the system for the 1060\,nm band raised two challenges. Firstly, the generation of stable, broadband, low-noise laser pulses at these wavelengths was not obvious. Secondly, supported by the mature telecommunication market, efficient dispersive elements such as long single mode fiber (SMF), dispersion compensating fiber (DCF) and chirped fiber Bragg gratings (CFBG) have been developed at 1550\,nm. These offered considerable amount of dispersion with low attenuation, which is not available at other wavelengths.

To tackle the first challenge, the use of incoherently broadened supercontinuum has been considered \cite{Moon2006}. However, the significant relative intensity noise (RIN) of these incoherently broadened spectra prevented high-quality OCT imaging. Recent research into supercontinuum dynamics in all-normal dispersion (ANDi) fiber demonstrated the production of temporally coherent light that is spectrally flat, with low RIN \cite{Genier2021,Sylvestre2021,Shreesha2021}. We previously reported the use ANDi supercontinuum combined with time-stretching to create a 40\,MHz swept-source for SS-OCT at 1060 nm \cite{grelet202240}. However, the stretcher then used was a 2.8\,km long SMF, leading to high attenuation and limited dispersion that gave low sensitivity and short axial range.
Recent development of long, broadband CFBG at 1050\,nm enabled stretching of pulses over 100\,ns which opened the door for an efficient supercontinuum-based time-stretched swept-source at 10\,MHz.

This letter details enhancements to the system described in \cite{grelet202240} including reduced laser noise, increased efficiency, lower losses, and a more compact stretcher. These improvements enable the demonstration of real-time \textit{in vivo} retinal imaging with improved sensitivity.

\section{Time-stretched supercontinuum swept-source} 
\begin{figure}[t]
\centering
\includegraphics[width=1\linewidth]{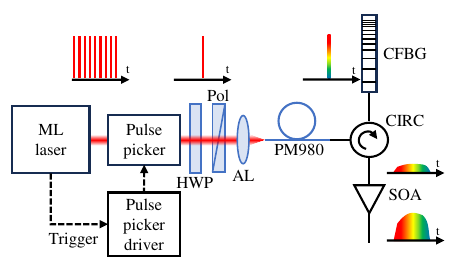}
\caption{Schematic of the time-stretched supercontinuum pulse swept-source structure and pulse evolution. ML laser, mode-locked laser; HWP, half-wave plate; Pol, polarizer; AL, aspheric lens; CIRC, circulator; CFBG, chirped fiber Bragg grating; SOA, semiconductor optical amplifier.}

\label{fig:SS_setup}
\end{figure}

Figure \ref{fig:SS_setup} presents the schematic of the time-stretched swept-source structure. A commercial mode-locked laser (NKT Photonics, Origami 10 HP) generated 220\,fs pulses at a repetition rate of 80\,MHz with a central wavelength at 1050\,nm. The repetition rate of the laser was then reduced to 10 MHz using a pulse-picker based on a commercial acousto-optic modulator (AOM) (G\&H, AOMO 3200-1113). A balance in the AOM parameters had to be found to maximize its raising speed, while preserving high diffraction efficiency. The AOM was triggered using the mode-locked laser internal photodiode signal and the pulse-picker gate time-delay was optimized to reduce the amplitude of the adjacent peaks that would otherwise add artefacts to the OCT signal.

\begin{figure}[t]
\centering
\includegraphics[width=1\linewidth]{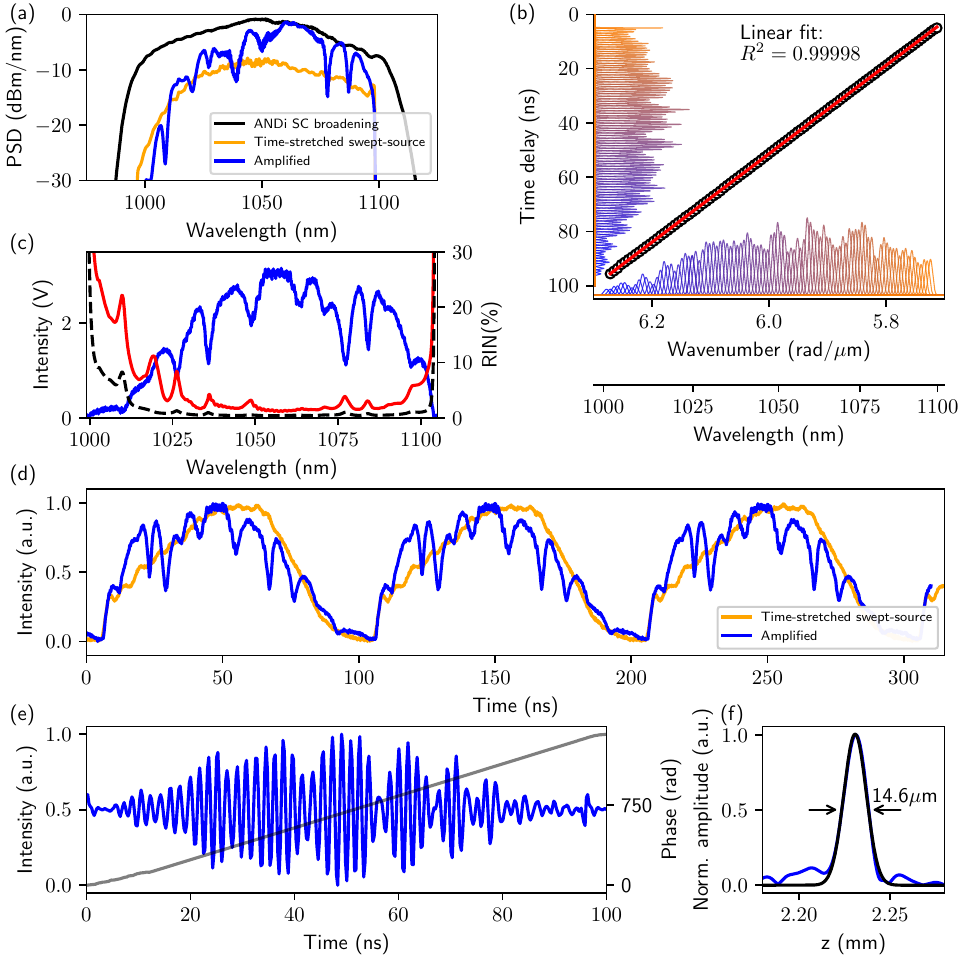}
\caption{(a) Spectrum at 3 points along the system. (b) Sweep linearity in wavenumber (wavelength axis non linear). (c) Intensity noise measurement (red), detector noise level (dashed black), swept-source spectrum (blue). (d) Time trace of the swept-source output. (e) Channeled spectrum at the interferometer output and sweep phase (black). (f) Axial resolution measurement with gaussian fit (black).}

\label{fig:SSperf}
\end{figure}

The beam was then coupled into a polarization maintaining (PM) fiber (Coherent, PM980). Due to the strong intensity of the electric field, the spectrum first broadens through self-phase modulation (SPM), and then through optical wave breaking (OWB) \cite{Heidt2011}. The short duration of the mode-locked laser pulses prevent noise from modulation instabilities and Raman scattering. To prevent polarized modulation instabilities (PMI) in the fiber that would seed intensity noise \cite{Gonzalo2018}, the polarization of the light was aligned on the fast axis of the PM fiber using a zero-order half-wave plate (Thorlabs, WPH10M-1053). A polarizer (Thorlabs, GTH10M) further ensured the linearity of the polarization state. To ensure the spectral broadening covering the time stretcher 100\,nm bandwidth, we used simulations based on the generalized Schrödinger equation (GNLSE) optimizing the fiber length and laser peak power. 5\,m of PM980 fiber were used, ensuring that SPM-generated spectral ripples are smoothed by the OWB. An average power of 102\,mW at 10\,MHz was used, reduced to 45\,mW (18\,kW peak power) at the fiber input due to attenuation from polarization optics and fiber coupling losses. The black curve of Fig. \ref{fig:SSperf} (a) presents the produced spectrum with a bandwidth of 60\,nm at FWHM, 110\,nm at -10\,dB. Compared to heavily structured spectrum of the anomalous dispersion supercontinuum reported in the literature, the ANDi supercontinuum exhibits a smoother spectrum with sharper edges \cite{Sylvestre2021}. 

The spectrally broadened pulses are then stretched in time using a broadband circulator (Opneti, HPCIR-3-1064-900-4-1-NE-1W) and a long CFBG (Proximion, CB-HB005-B0930FA). The grating had a transmission bandwidth $\Delta\lambda$ of 100 nm centered at $\lambda_0$=1050\,nm, and a dispersion parameter D=930\,ps/nm. This stretches the pulse to $\Delta t$=D$\cdot\Delta\lambda$=93\,ns, where the steep spectral edges produced by OWB translates to a duty cycle of 93\% in the time domain as shown in Fig.\ref{fig:SSperf} (d). The sweep linearity was assessed by filtering part of the spectrum with a monochromator, measuring the filtered spectrum with an optical spectrum analyzer (Yokogawa AQ6373B), and determining the time delay to a fixed reference using a 5\,GHz photodiode (Thorlabs DET08CFC/M) and a 12\,GHz sampling oscilloscope (Picotech Picoscope 9200).
A linear regression applied to the data, as presented in Fig.\ref{fig:SSperf} (b), demonstrated a high sweep linearity in wavenumber with a coefficient of determination $R^2$ of 0.99998. Thus, data resampling before FFT processing is not required in this system.

A key advantage of the long CFBG is its low attenuation-to-dispersion ratio. The combined attenuation of the circulator and CFBG was measured at 7.1\,dB, resulting in 8.7\,mW at the swept-source output. In comparison, achieving the same dispersion with over 18\,km of step-index fiber (Coherent 980XP) would result in more than 46\,dB attenuation.
Although this power suffices for OCT imaging of highly reflective samples, optical amplification proved necessary for SS-OCT of biomedical samples exhibiting high scattering. A broadband semiconductor optical amplifier (SOA) (Innolume, SOA-1020-110-HI-27dB) was used to enhance the swept-source average power to 37\,mW.

We notice a spectral reshaping by the stretcher with higher attenuation of the low wavelengths, reducing the bandwidth to 55\,nm at FWHM and 100\,nm at -10\,dB. Significant spectral modifications were observed after the SOA, reducing the bandwidth to 35\,nm at FWHM and 90\,nm at -10\,dB. This is due to polarization mode dispersion (PMD) in the circulator and long CFBG, revealed by the SOA gain, which is exclusively for the S linear polarization. PMD compensation across the entire bandwidth with a polarization controller is impractical due to wavelength-dependent birefringence in the CFBG. Future experiments could address this by writing the CFBG on a PM fiber.

ANDi supercontinuum dynamics are recommended for low-noise broadband generation. Their recent use in time-stretched swept-source systems warrants investigation of noise transfer. Fig. \ref{fig:SSperf} (c) presents a measurement of the RIN of the swept-source output. 10,000 consecutive sweeps were recorded using a 2.5\,GHz photodiode (Thorlabs, PDB482C-AC) and a 20\,GHz oscilloscope (Teledyne Lecroy, WaveMaster 820Zi-B). The RIN was calculated using \cite{Smith2022}:
\begin{equation}
    RIN(\lambda)=\frac{\sigma(\lambda)}{\mu(\lambda)}
\end{equation}
The RIN measured is as low as 1.2\%. We note that it remains slightly above the photodetector noise floor, which may result from amplified stimulated emission (ASE) in the SOA.

\begin{figure}[ht]
\centering
\includegraphics[width=1\linewidth]{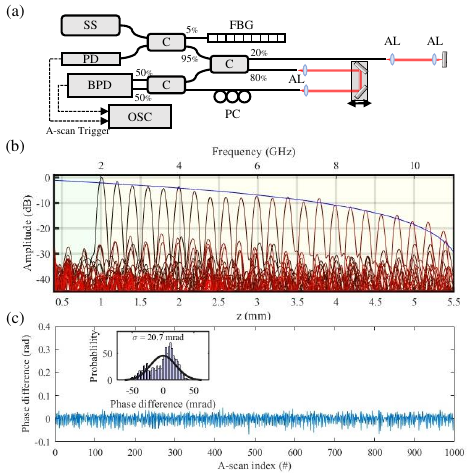}
\caption{(a) Schematic of the SS-OCT set-up. SS, swept-source; C, fiber coupler; FBG, fiber Bragg grating; AL, aspheric lens; PC, polarization controller; PD, photodetector; BPD, balanced photodetector; OSC, Oscilloscope. (b) Roll-off measurement. The yellow and the green areas respectively represent the OCT axial range achievable with the 20 GHz OSC used for characterization, and with the 2 GHz DAC used for the retinal imaging. (c) Phase noise measurement.}

\label{fig:SS-OCT setup}
\end{figure}

\section{10 MHz Swept-Source OCT}
Figure \ref{fig:SS-OCT setup} (a) presents the structure of the OCT system used to characterize the OCT performance using the swept-source. This system adopts an asymmetric design to optimize light collection from the sample, a critical aspect when imaging low reflectivity samples. An InGaAs 22\,GHz balanced photodetector (Optilab, BPR-23-M) and a 20\,GHz oscilloscope digitized the interference signal. A 400\,MHz high-pass filter removes the DC signal.
To limit dispersion imbalance, the reference and sample arms lengths were precisely matched. At the system input, a 95/05 fiber coupler and a fiber Bragg grating with a 0.1\,nm bandwidth centered at 1090\,nm reflect a limited part of the spectrum. The reflected light was detected with a 5\,GHz photodiode and used as a trigger for the OCT system.

The interference signal was processed using the Complex Master Slave (CMS) method \cite{Rivet2016}. Thirty interference signals were recorded at various optical path differences to accurately measure the static and depth-dependent phase. From these, 10,000\,eigenvectors were synthesized, each corresponding to a specific depth with 0.55\,µm spacing. While the PDM-based structure reduced the PSD at certain wavelengths, it did not hinder phase extraction of the masks. A Hamming apodization was applied to the synthetic masks to limit side lobes in the A-scan. Figure \ref{fig:SSperf} (e) and (f) show a recorded channeled spectrum and an A-scan, respectively, with an axial resolution in air of $\delta z$=14.6\,µm, consistent with the theoretical resolution.

A roll-off measurement is presented in Fig.\ref{fig:SS-OCT setup} (b), with 10\,dB axial range of 3.56\,mm.
Phase stability is crucial for many applications. Using a common path interferometer, we measured a phase noise of 20.7\,mrad, consistent with other reports on high sweep rates \cite{Wei2015}, as presented in Fig. \ref{fig:SS-OCT setup}\,(c). Long-term phase stability was assessed over two hours by periodically generating Ascans with previously recorded masks. The Ascans' FWHM varied by less than 5\%, with no significant thermal dependency observed in the phase sweep.

\begin{figure}[ht]
\centering
\includegraphics[width=1\linewidth]{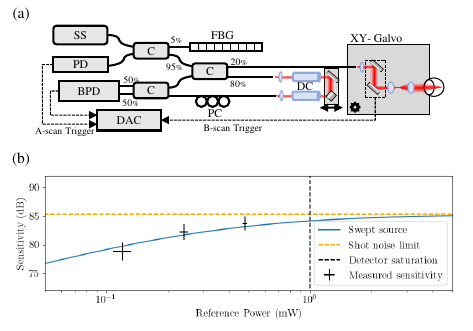}
\caption{(a) OCT set-up used for retinal imaging. DC, dispersion compensation glass; DAC, digital-analog converter. (b) Sensitivity of the SS-OCT system (measured and theoretical).}
\label{fig:OCT_retinalImaging}
\end{figure}
To demonstrate retinal imaging, the imaging system was modified as presented in Fig. \ref{fig:OCT_retinalImaging} (a). The fiber output of the sample arm was mounted on a motorized stage, enabling pupil tracking and control of the sample arm length by adjusting the focal point. The beam was deflected by galvanometer scanner mirrors (Cambridge Technology, 6210H). A telescope was integrated to maximize lateral resolution, with dispersion compensation glass rods in the reference arm. The lateral resolution, measured using a 1951 USAF target, was $\sim$ 19.69\,µm (group\,4,\,element\,5).

Real-time visualization is critical for eye positioning control. Therefore, a 12-bit waveform digitizer (AlazarTech, ATS9373) with a 4\,GS/s acquisition rate was incorporated for live processing. A balanced photodetector (Thorlabs, PDB482C-AC) with smaller bandwidth but better noise performance was used. A new calibration was performed using a model eye composed of a 19\,mm aspheric lens (Thorlabs, AC127-019-B-ML) and a flat silver mirror. The sample power was 3.9\,mW.

Figure \ref{fig:OCT_retinalImaging} (b) shows the sensitivity measurement for multiple reference power, reaching up to 84 dB which corresponds to the shot-noise limited sensitivity, calculated as \cite{DeBoer2017}:
\begin{equation}
    S_{shot-noise}=10\ log_{10}\left(\frac{\rho\  P_{sample}}{e\ f_{sweep}}\right)-IL
\end{equation}
with $\rho$\,=\,0.72\,A/W the photodiode responsivity at 1060\,nm, $P_{sample}$ the optical power incident on the sample, e the elementary charge of the electron, $f_{sweep}$ the sweeping rate, and IL the loss on the way back to the photodetector.

\begin{figure}[ht]
\centering
\includegraphics[width=1\linewidth]{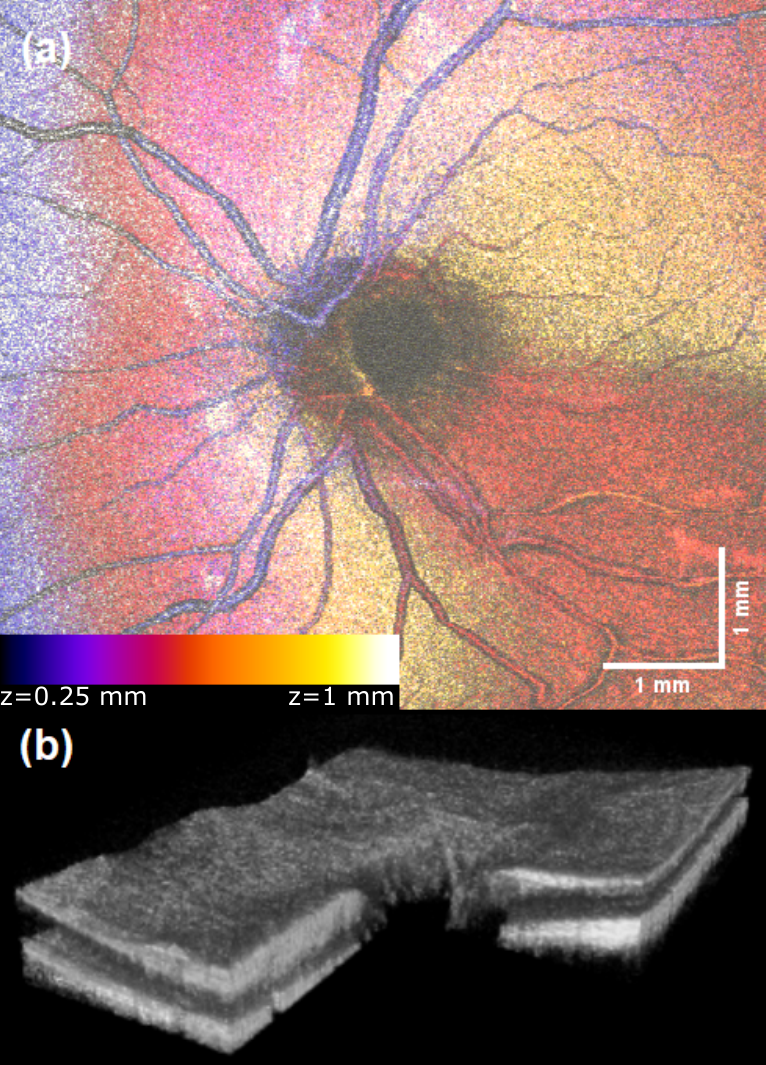}
\caption{(a) Assembly of 500 \textit{en face} views of a human retina with color coded depth. (b) Volume of the same retina with a cut showing B-scan of the optic nerve.}
\label{fig:Retina_enface&volume}
\end{figure}

Figure \ref{fig:Retina_enface&volume} demonstrates \textit{in vivo} imaging of a human retina with an \textit{en face} view and a volume view. The area imaged is 6.75\,mm by 8\,mm. The \textit{x}-axis galvoscanner is driven with a 1.5 kHz triangular signal. The data are recorded during the forward scan and processed during the backward scan, leading to a B-scan acquisition time of 0.33\,ms. Each B-scan contains 2200\,A-scans, and each volume consists of 500\,B-scans. This leads to a volume rate of 4.5\,Hz. The optic nerve is well-defined, and blood vessels are clearly distinguishable.

A faster volume acquisition rate can be achieved with a faster scanner on the \textit{x}-axis, such as a resonant scanner. Further improvement of the OCT system could also exploit the high B-scan rate for OCTA imaging. Finally, using a 10 MHz pump laser and all-PM components (circulator and CFBG) would simplify the swept-source and improve its performances.

\section{Conclusion}
In this letter, a swept-source for shot-noise limited multi-MHz OCT applications is presented, specifically geared towards high-speed retinal imaging at 1060 nm. It is based on the combination of ANDi supercontinuum dynamics that provides broad, smooth and low noise spectrum with time-stretching. A high tuning speed of 10\,MHz is achieved with a 93\% duty cycle, a sweep linearity higher than $R^2$=0.99998, and low intensity noise. 
The capabilities of this source for SS-OCT were demonstrated, and high-speed \textit{in vivo} human retina imaging was showcased. 
This method opens a promising path toward high acquisition speed OCT, reducing the effects of object movement, as required for OCT of the eye, surgery monitoring and OCTA. More work would be required to further optimize the pump laser.

\section*{Acknowledgments}
This project has received funding from the European Union's Horizon 2020 research and innovation programme Marie Sklodowska-Curie grant agreement No 860807. AP also acknowledges the NIHR BRC4-05-RB413-302, Imaging, Visual Assessment \& Digital Innovation, at Moorfields Eye Hospital NHS FT and UCL Institute of Ophthalmology the BBSRC BB/S016643/1 and BB/X003744/1, the NIHR 202879, the MRC Impact Accelerator Account, project 18557, the DPFS MRC 662915 to UoNottingham and UoKent and the EPSRC, 26049, FoVEnOCT, EP/X000125/1.

\section*{Disclosures}
Adrian Podoleanu is coinventor on patents in the name of the University of Kent.

\bibliographystyle{IEEEtran}
\bibliography{bibtex/bib/sample.bib}

\begin{thebibliography}{10}
\providecommand{\url}[1]{#1}
\csname url@samestyle\endcsname
\providecommand{\newblock}{\relax}
\providecommand{\bibinfo}[2]{#2}
\providecommand{\BIBentrySTDinterwordspacing}{\spaceskip=0pt\relax}
\providecommand{\BIBentryALTinterwordstretchfactor}{4}
\providecommand{\BIBentryALTinterwordspacing}{\spaceskip=\fontdimen2\font plus
\BIBentryALTinterwordstretchfactor\fontdimen3\font minus \fontdimen4\font\relax}
\providecommand{\BIBforeignlanguage}[2]{{%
\expandafter\ifx\csname l@#1\endcsname\relax
\typeout{** WARNING: IEEEtran.bst: No hyphenation pattern has been}%
\typeout{** loaded for the language `#1'. Using the pattern for}%
\typeout{** the default language instead.}%
\else
\language=\csname l@#1\endcsname
\fi
#2}}
\providecommand{\BIBdecl}{\relax}
\BIBdecl

\bibitem{bussel2014oct}
I.~I. Bussel, G.~Wollstein, and J.~S. Schuman, ``Oct for glaucoma diagnosis, screening and detection of glaucoma progression,'' \emph{British Journal of Ophthalmology}, vol.~98, no. Suppl 2, pp. ii15--ii19, 2014.

\bibitem{schlegl2018fully}
T.~Schlegl, S.~M. Waldstein, H.~Bogunovic, F.~Endstra{\ss}er, A.~Sadeghipour, A.-M. Philip, D.~Podkowinski, B.~S. Gerendas, G.~Langs, and U.~Schmidt-Erfurth, ``Fully automated detection and quantification of macular fluid in oct using deep learning,'' \emph{Ophthalmology}, vol. 125, no.~4, pp. 549--558, 2018.

\bibitem{Draxinger:24}
\BIBentryALTinterwordspacing
W.~Draxinger, N.~Detrez, P.~Strenge, V.~Danicke, D.~Theisen-Kunde, L.~Sch\"{u}tzeck, S.~Spahr-Hess, P.~Kuppler, J.~Kren, W.~Wieser, M.~M. Bonsanto, R.~Brinkmann, and R.~Huber, ``Microscope integrated mhz optical coherence tomography system for neurosurgery: development and clinical in-vivo imaging,'' \emph{Biomed. Opt. Express}, vol.~15, no.~10, pp. 5960--5979, Oct 2024. [Online]. Available: \url{https://opg.optica.org/boe/abstract.cfm?URI=boe-15-10-5960}
\BIBentrySTDinterwordspacing

\bibitem{Wei2019}
X.~Wei, T.~T. Hormel, S.~Pi, Y.~Guo, Y.~Jian, and Y.~Jia, ``High dynamic range optical coherence tomography angiography (hdr-octa),'' \emph{Biomedical Optics Express}, vol.~10, p. 3560, 7 2019.

\bibitem{Kim2020}
T.~S. Kim, J.~Y. Joo, I.~Shin, P.~Shin, W.~J. Kang, B.~J. Vakoc, and W.~Y. Oh, ``9.4 mhz a-line rate optical coherence tomography at 1300 nm using a wavelength-swept laser based on stretched-pulse active mode-locking,'' \emph{Scientific Reports}, vol.~10, 12 2020.

\bibitem{Klein2017}
T.~Klein and R.~Huber, ``High-speed oct light sources and systems [invited],'' \emph{Biomedical Optics Express}, vol.~8, p. 828, 2 2017.

\bibitem{Potsaid2012}
B.~Potsaid, V.~Jayaraman, J.~G. Fujimoto, J.~Jiang, P.~J.~S. Heim, and A.~E. Cable, ``Mems tunable vcsel light source for ultrahigh speed 60khz - 1mhz axial scan rate and long range centimeter class oct imaging,'' \emph{SPIE BiOS XII}, vol. 8213, 2 2012.

\bibitem{huber2005amplified}
R.~Huber, M.~Wojtkowski, K.~Taira, J.~G. Fujimoto, and K.~Hsu, ``Amplified, frequency swept lasers for frequency domain reflectometry and oct imaging: design and scaling principles,'' \emph{Optics Express}, vol.~13, no.~9, pp. 3513--3528, 2005.

\bibitem{Akkaya2022}
I.~Akkaya and S.~Tozburun, ``A stretched-pulse mode-locked (spml) wavelength-swept laser source at 1.06 µm,'' \emph{SPIE-Intl Soc.Opt.Eng.}, p.~45, 3 2022.

\bibitem{Huo2015}
T.~Huo, C.~Wang, X.~Zhang, T.~Chen, W.~Liao, W.~Zhang, S.~Ai, J.-C. Hsieh, and P.~Xue, ``Ultrahigh-speed optical coherence tomography utilizing all-optical 40 mhz swept-source,'' \emph{Journal of Biomedical Optics}, vol.~20, p. 030503, 3 2015.

\bibitem{Huang2020}
D.~Huang, F.~Li, Z.~He, Z.~Cheng, C.~Shang, and P.~K.~A. Wai, ``400 mhz ultrafast optical coherence tomography,'' \emph{Optics Letters}, vol.~45, p. 6675, 12 2020.

\bibitem{Moon2006}
S.~Moon and D.~Y. Kim, ``Ultra-high-speed optical coherence tomography with a stretched pulse supercontinuum source,'' \emph{Optics Express}, vol.~14, pp. 1178--1181, 2006.

\bibitem{Genier2021}
E.~Genier, S.~Grelet, R.~D. Engelsholm, P.~Bowen, P.~M. Moselund, O.~Bang, J.~M. Dudley, and T.~Sylvestre, ``Ultra-flat, low-noise, and linearly polarized fiber supercontinuum source covering 670–1390 nm,'' \emph{Optics Letters}, vol.~46, p. 1820, 4 2021.

\bibitem{Sylvestre2021}
\BIBentryALTinterwordspacing
T.~Sylvestre, E.~Genier, A.~N. Ghosh, P.~Bowen, G.~Genty, J.~Troles, A.~Mussot, A.~C. Peacock, M.~Klimczak, A.~M. Heidt, J.~C. Travers, O.~Bang, and J.~M. Dudley, ``Recent advances in supercontinuum generation in specialty optical fibers [invited],'' \emph{Journal of the Optical Society of America B}, 11 2021. [Online]. Available: \url{http://arxiv.org/abs/2111.03719 http://dx.doi.org/10.1364/JOSAB.439330}
\BIBentrySTDinterwordspacing

\bibitem{Shreesha2021}
S.~R.~D. S, M.~Jensen, L.~Grüner-Nielsen, J.~T. Olsen, P.~Heiduschka, B.~Kemper, J.~Schnekenburger, M.~Glud, M.~Mogensen, N.~M. Israelsen, and O.~Bang, ``Shot-noise limited, supercontinuum-based optical coherence tomography,'' \emph{Light: Science and Applications}, vol.~10, 12 2021.

\bibitem{grelet202240}
S.~Grelet, A.~M. Jim{\'e}nez, R.~D. Engelsholm, P.~B. Montague, and A.~Podoleanu, ``40 mhz swept-source optical coherence tomography at 1060 nm using a time-stretch and supercontinuum spectral broadening dynamics,'' \emph{IEEE Photonics Journal}, vol.~14, no.~6, pp. 1--6, 2022.

\bibitem{Heidt2011}
\BIBentryALTinterwordspacing
A.~M. Heidt, A.~Hartung, G.~W. Bosman, P.~Krok, E.~G. Rohwer, H.~Schwoerer, and H.~Bartelt, ``Coherent octave spanning near-infrared and visible supercontinuum generation in all-normal dispersion photonic crystal fibers,'' \emph{Optics Express}, vol.~19, pp. 3775--3787, 2011. [Online]. Available: \url{http://www.nktphotonics.com}
\BIBentrySTDinterwordspacing

\bibitem{Gonzalo2018}
I.~B. Gonzalo, R.~D. Engelsholm, M.~P. Sørensen, and O.~Bang, ``Polarization noise places severe constraints on coherence of all-normal dispersion femtosecond supercontinuum generation,'' \emph{Scientific Reports}, vol.~8, 12 2018.

\bibitem{Smith2022}
C.~R. Smith, R.~D. Engelsholm, and O.~Bang, ``Pulse-to-pulse relative intensity noise measurements for ultrafast lasers,'' \emph{Optics Express}, vol.~30, p. 8136, 2 2022.

\bibitem{Rivet2016}
S.~Rivet, M.~Maria, A.~Bradu, T.~Feuchter, L.~Leick, and A.~Podoleanu, ``Complex master slave interferometry,'' \emph{Optics Express}, vol.~24, p. 2885, 2 2016.

\bibitem{Wei2015}
X.~Wei, A.~K.~S. Lau, Y.~Xu, K.~K. Tsia, and K.~K.~Y. Wong, ``28 mhz swept source at 10 um for ultrafast quantitative phase imaging,'' \emph{Biomedical Optics Express}, vol.~6, p. 3855, 10 2015.

\bibitem{DeBoer2017}
J.~F. de~Boer, R.~Leitgeb, and M.~Wojtkowski, ``Twenty-five years of optical coherence tomography: the paradigm shift in sensitivity and speed provided by fourier domain oct [invited],'' \emph{Biomedical Optics Express}, vol.~8, p. 3248, 7 2017.

\end{thebibliography}


 




\vfill

\end{document}